\documentstyle[12pt]{article}
\author{Hans-J\"urgen Schmidt }
\title{A new duality transformation for 
fourth--order gravity}
\date{}
\begin{document}
\maketitle

\centerline{
 University Potsdam, Institute for Mathematics, Cosmology group}
\centerline{
      D-14415 POTSDAM, PF 601553, Am Neuen Palais 10, Germany}
\centerline{e-mail \  \  \   hjschmi@rz.uni-potsdam.de }

\begin{abstract} 

We prove that for non--linear $L=L(R)$, $G=dL/dR \ne 0$ the
Lagrangians
$L$ and $ \hat L ( \hat R)$ with
$\hat L  =  2R/G\sp 3  -  3 L/G \sp 4 $, 
 $\hat g_{ij} = G \sp 2 \, g_{ij}$ and
$\hat R =  3R/G\sp 2  -  4 L/G \sp 3 $ 
give conformally equivalent fourth--order field equations being
dual to each  other. The proof represents a new
application of the fact that the
operator $\Box - \frac{R}{6}$ is conformally invariant.

\medskip

\end{abstract}

Gen. Rel. Grav. in print; 

KEY: Conformal relations in fourth--order gravity 

\medskip

\section{Introduction}

Higher--order theories, especially fourth--order gravity
theories, are subject to conflicting facts: On the one hand, they
appear quite naturally from generally accepted principles; on the
other hand, they are unstable and so, 
 they should be considered unphysical.

Nevertheless, from time to time one can find  a 
statement like: ''Whichever turns out to be the real 
theory of gravitation,
 the corresponding low--energy effective Lagrangian will  
probably contain higher derivative terms.'' [1], so 
it makes sense to elucidate the structure of such theories
without specifying a concrete physical context. 

\medskip

In sct. 2, we shortly review some classes of conformally
related theories, especially, to show the difference to
our new approach of sct. 3. 

\medskip

     Sct. 3 deals with fourth--order gravity following from a
non--linear Lagrangian $L(R)$. The conformal equivalence of these
theories to theories of other types is widely known, but the
conformal equivalence of these theories to theories of the same
type but with different Lagrangian is much less known. We
fill this gap by proving a duality theorem between pairs of such
fourth--order theories. 

\medskip


\section{Conformally related theories}

In Rainich (1925, ref. [2]) the electromagnetic field was
calculated from the curvature tensor. This was  
cited in Kucha\v r (1963, ref. [3]) as example for the
geometrization programme; in [3]  
on meson fields $\psi$ (now called scalar fields), Kucha\v r  
gives a kind of geometrization by using a relation between $\psi$
and $R$, then he gets the equation 
$$\Box R - \frac{k^2}{2} R=0$$
which is of fourth order in the metric. This is
the trace of 
fourth--order gravity as we are dealt with, but he did not
deduce it from a curvature squared action.

\setcounter{equation}{0}

In  Bekenstein  (1974, ref. [4]) 
the conformal transformation from Einstein's theory with a
minimally coupled ($\phi $) to a conformally coupled ($\psi$) 
scalar field is proven where additional conformally invariant  
matter (radiation) is allowed. For $8\pi G=1$ one has 
$$\psi \ = \ \sqrt 6 \tanh ( \phi /\sqrt 6 )$$
 If radiation is absent and $\phi \ne 0$ 
 then it works also with "coth" instead of
"tanh". This is reformulated in his theorem 2:
If $g_{ij}$ and $\psi$ form an Einstein--conformal scalar
solution
with $\psi^2 \ne 6$, 
then $\hat g_{ij} = \frac{1}{6} \psi^2  g_{ij}$ and 
$\hat \psi = 6/ \psi$ form a second one. One can see that this is
a dual map because by applying the operator $\ \hat{} \ $ twice,
the original solution is re--obtained.

Let us comment this theorem 2: For the conformal scalar field one
has the effective gravitational constant $G_{eff}$ defined by  
$$ \frac{1}{8 \pi G_{eff}} \ = \ 1 \  -  \  \frac{\psi^2}{6}$$ A
positive value $G_{eff}$ implies a negative value 
$\hat G_{eff}$
 and vice versa. 
 By changing the overall sign of the Lagrangian
one can achieve a positive effective gravitational  constant at
the price of the scalar field becoming a ghost (wrong sign in
front of the kinetic term). So, Bekenstein has given a conformal
transformation from Einstein's theory with a conformally coupled
ordinary scalar field to Einstein's theory  with a conformally
coupled ghost. From this property one can see that this duality
relation is different from the duality theorem to be deduced 
at in sct. 3, because there 
are no ghosts at all.

Later but independently of [4] the 
conformal equivalence between minimally and conformally coupled
scalar fields with $G_{eff} > 0$ was generalized in [5] by the
inclusion of several self--interaction terms. Cf. also ref. [6]
for further generalizations to arbitrary coupling parameter  
$\xi$ and higher dimensions.

The conformal transformation from fourth--order gravity to
Einstein's theory with a minimally coupled scalar field was
deduced in several steps:  Bicknell (1974, ref. [7]) found the
  transformation for $L= \frac{1}{2} R^2$; the 
conformal factor is 
$\vert R \vert $, and
after the transformation, 
which is valid for $R \ne 0$, 
 one gets Einstein's theory with
non--vanishing $\Lambda$-term and a massless minimally coupled
scalar field as source.

Similar results for $L(R)$ and conformal factor
$\vert \frac{dL}{dR} \vert $
 have been deduced in [8], where 
$ \frac{dL}{dR}  \ne 0 $ and
$$ \frac{d^2L}{dR^2} \  \ne \  0 $$
have to be fulfilled. The second paper
of ref. [8] generalized the conformal transformation to more
general tranformations of the metric.  
 Later on and in a different context (metric--affine
theories) such transformations have been considered also by
Jakubiec and Kijowski [9]. 
 Ref. [10] generalized these conformal transformations           
by the inclusion of non--minimally coupled scalar fields. 
 In [6] it is noted that it may happen
 that one singular and another non--singular  model may 
be conformally related. 
  Magnano and  Soko\l owski  [11] discuss which of the
conformally related frames can be considered to be physical.

\section{A duality theorem}

Let us now deduce the duality theorem announced in the
introduction which shall close a gap in the set of  the        
aforementioned results. Two predecessors exist already: 

\medskip

  Buchdahl (1978, ref. [12]) showed: For $L = \frac{1}{2}R^2$ 
 the conformal
factor $R^2$ (if it is $\ne 0$) transforms solutions to solutions
and represents a dual map in the set of solutions. (Another
conformal factor than in ref. [7], cf. sct. 2).

\medskip

Ref. [13] generalizes this dual map to
other non--linear Lagrangians
 $L(R) = \frac{1}{k+1} R^{k+1}$ , \  
 (where $k \ne -1, \, 0$), the conformal factor being
$R^{2k}$, and $R$ is supposed to be different from zero. 
(Again, this conformal factor is the square of that  conformal
factor which is necessary to transform to Einstein's theory with
a minimally coupled scalar field.)

\subsection{The general 4--dimensional case}

\medskip

Let us now start with the key element of the deduction. For a
metric $g_{ij}$ and a scalar $G\ne 0$ we define the conformally
related metric 
\begin{equation}
\hat g_{ij}=G^{2}g_{ij}                   
\end{equation}
Our strategy is to develop the hat $ \wedge $
 to a duality operator: 
Every valid statement shall remain correct if all hats are
removed and all formerly unhatted quantities get a hat.
For eq. (1) this means $g_{ij}=\hat G^{2} \ \hat g_{ij}$. So
duality requires 
\begin{equation}
    G \  \hat G \ = \ 1
\end{equation}
implying $\hat G \ne 0$.
For 
$$\Box_{c}\equiv \Box - \frac{R}{6}$$
 the validity of
\begin{equation}
G^{3} \ \hat{} \, \Box_{c}  \ = \ \Box_c G
\end{equation}
reflects the conformal invariance of the operator $\Box_c$ if
applied to any scalar $\chi$. 
Now we apply eq. (3) to the constant
scalar $ \chi =-6$ and get
\begin{equation}
 G^3 \hat R \ = \  GR - 6\Box  G
\end{equation}
Duality implies
\begin{equation}
\hat G^{3} R \ = \ \hat G \hat R - 6 \ 
 \hat{} \, \Box  \hat G        
\end{equation}
Now we are prepared to consider a gravitational Lagrangian
$L=L(R)$ where $L$ is a smooth function (or, at least
 $C^3$--differentiable). We define $G=\frac{dL}{dR}$ and 
$H=\frac{d^{2}L}{dR^2}$ and restrict to an $R$-interval where
$GH\ne0$.
Then $L$ gives rise to a fourth-order field equation. We 
decompose this equation into the trace
\begin{equation}
3\Box  G \ = \  2 L - G R    
\end{equation}
and the trace-free part. The latter is equivalent to require that
the trace-free part of the tensor $G R_{ij}-G_{;ij}$
vanishes.
We insert eq. (6) into eq. (4) and get
\begin{equation}
\hat R = \frac {3R}{G^2}- \frac {4L}{G^3}    
\end{equation}
The dual to eq. (7) reads
\begin{equation}
R=\frac {3 \hat R}{\hat G ^2} - \frac {4 \hat L} {\hat G ^3}    
\end{equation}
For $\hat R$ we insert the expression (7), for $\hat G$ we use
eq. (2), and then we can solve eq. (8) for $\hat L$ as
follows
\begin{equation}
\hat L = \frac {2 R}{G^3} -  \frac {3 L}{G^4}   
\end{equation}
Applying $\frac {d}{dR}$ to eq. (7) we get
\begin{equation}
\frac {d \hat R}{dR} = \frac {6H}{G^4} (2L - GR) -
 \frac {1}{G^2}    
\end{equation}
with the consequence that $\frac {d \hat R}{dR} \ne 0$
if and only if
\begin{equation}
G^2 \ne 6H (2L-GR)    
\end{equation}
Let (11) be fulfilled in the following, then eq. (7) can be
locally inverted as $R= R (\hat R)$ and with (9) we get 
$\hat L = \hat L (\hat R)$.
It is useful to make the following consistency test: 
Calculate
$$
\hat G = \frac{d \hat L}{d \hat R} = 
\frac{d \hat L}{d R} 
\cdot \left(  \frac{d \hat R}{dR}  \right) ^{-1}
$$
via eqs. (9), (10); one gets $\hat G = \frac {1}{G}$ 
consistent with eq. (2). The analogous consistency takes place
with the fourth-order field equation. This proves the following

\medskip
\noindent
{\bf Theorem:}
 Let $g_{ij}$ be a solution of the field equation
following from $L(R)$ then $\hat g_{ij}$ is a solution for 
$\hat L (\hat R)$. 

\medskip

The duality theorem deduced above is a method 
to construct new solutions of fourth--order gravity from known
solutions of a (possibly other) fourth--order theory. 

This theorem is most powerful if applied to solutions with 
non-constant value $R$; the reason is obvious: On the one hand,
the solutions with constant $R$ are identical to solutions 
of Einstein's 
vacuum equation with suitably chosen cosmological term,
and, on the other hand, the conformal factor is constant 
for this case. 

If we replace $L(R)$ by $c \cdot L(R)$ then $\hat L ( \hat R)$
is changed to  $\hat L ( \hat R/c^2)/c^3$. So, up to 
a scale-transformation, nothing is changed. 

$L$ and $\hat L$ represent the same function if and only if the
corresponding potential $V(\Phi)$ is an even function
in $\Phi$ (in that conformal picture where Einstein's theory
with the minimally coupled scalar field $\Phi$ and 
potential $V$ is applied, cf. refs. [8]).

\subsection{Special examples}

\medskip

For $L=\frac {1}{2} R^2$ we get $G=R$, 
$\hat g_{ij} = R^2 g_{ij}$, i.e., only the 
range $R \ne 0$ is allowed. The trace  
eq. (6) reduces to $\Box R=0$; and eq. (4) can be written as
$$
\hat R = \frac {1}{R} - \frac{6}{R} \Box R
$$
an identity which might be useful in another context, too; it
has the following corollary:  Let $R \ne 0$ and
 $\hat g_{ij} = R^2 g_{ij}$.
Then the following 3 equations are equivalent:
$$\Box R =0,  \qquad \hat{} \, \Box \hat R = 0 , 
 \qquad  R \hat R = 1
$$
Now let $L=\frac{1}{k+1} \vert R \vert ^{k+1}$ with $R\ne 0$
 and 
$k \ne -1,0$. $k=1$ leads to the above case. We have 
$G= \pm \vert R \vert ^k$ where the lower sign 
corresponds to the
case $R<0$. We get
$$\hat g_{ij} \ = \ \vert R \vert ^{2k} g_{ij}$$ 
and 
$$\hat R \ = \ \frac{3k-1}{k+1}  \cdot 
\frac{R}{\vert R \vert ^{2k}}$$
Requirement (11) reads $1\ne 6 k \cdot \frac {1-k}{1+k}$
and implies $k \ne \frac {1}{3}, \frac {1}{2}$.
We get 
$$\hat L \ = \ \frac{2k-1}{k+1} \cdot 
\frac{\vert R \vert}{\vert R \vert^{3k}}
\ = \ \hat c \cdot \vert \hat R \vert ^{\hat k + 1 }
$$
with $\hat k = \frac{1}{2- 1/k}$ 
 and a certain
 $\hat c (k) \ne 0$. 
The restrictions  $k \ne -1,0$ are immediately clear because 
$L(R)$ has to be nonlinear in $R$. To elucidate the restrictions
$k \ne \frac {1}{3}, \frac {1}{2}$ we rewrite eqs. (7)/(9) as
follows which is valid in the range $R>0$
($R<0$ is quite similar to deal with)
\begin{equation}
\hat R \ = \ \frac {3R^{7/3}}{G^3} \ \frac{d}{dR} 
\left(\frac{L}{R^{4/3}} \right)
\end{equation}
and
\begin{equation}
\hat L \ = \ \frac {2 R^{5/2}}{G^4} \ \frac{d}{dR} 
\left( \frac{L}{R^{3/2}} \right)  
\end{equation}

So, for $k=\frac{1}{3}$ we get 
$\hat R \equiv 0$, for $k=\frac{1}{2}$ we get
 $\hat L \equiv 0$.

\medskip

A new example reads as follows: Let
 $L=\frac {1}{4} (3R^{4/3} - 1)$,  
and let $R>0$. We get $G=R^{1/3}$, $\hat R=\frac{1}{R}$, and 
$\hat L (\hat R)$ is the same function as $L(R)$. So it holds: If
$g_{ij}$ is a solution of the corresponding field equation with
$R>0$ then $R^{2/3} g_{ij}$ is a solution, too. (In the
conformal picture with Einstein's theory this example
corresponds to a potential   
$V(\Phi)$ which is a multiple of $\cosh(\sqrt{8/3} \Phi)$.)

\subsection{Generalization to higher order and
higher dimension}

In [14], the conformal invariance of the operator 
${\bf D }$ defined by
$$ {\bf D} = \Box^2 + 2 R^{ij} \nabla_i\nabla_j 
 - \frac{2}{3} R \Box + \frac{1}{3} R^{;i} \nabla_i
$$
in four dimensions was shown. More detailed: Let 
$\hat g_{ij} =G^2 g_{ij}$ then 
$$G^4 \ \hat {\bf D } \ = \ {\bf D} $$
if applied to arbitrary scalars. It would be 
interesting to look whether this identity 
has similar consequences as eq. (3). 

\bigskip

Up to now, we have considered the 
4--dimensional case only. Let us give the corresponding result
for an arbitrary dimension $n \ge 3$. We get 
$$\Box_{c}\equiv \Box - \xi R$$
where $\xi = \frac{n-2}{4(n-1)}$, and we follow the
route sketched in ref. [15], sct. 3 and in Maeda ref. [10]. 
We replace eq. (1) by 
\begin{equation}
\hat g_{ij}=G^{4/(n-2)}g_{ij}                   
\end{equation}
for any non--vanishing scalar $G$. Without loss 
of generality we may assume $G > 0$ 
(for, otherwise, we might replace $L$ by $-L$ ), 
 and
 we 
keep $G=\frac{dL}{dR}$,  
$H=\frac{d^{2}L}{dR^2} \ne 0$, $\hat G = 1/G$. 
The 
trace-free part is again equivalent to require that
the trace-free part of the tensor $G R_{ij}-G_{;ij}$
vanishes. However, the trace of the field equation
following from $L(R)$ changes
and is equivalent to 
\begin{equation}
(n-1) \Box  G \ = \  \frac{n}{2} L - G R    
\end{equation}
in place  of eq. (6). Eq. (3) has 
to be replaced by (cf. e. g. [6] to see how to 
calculate the necessary powers of $G$) 
\begin{equation}
G^{(n+2)/(n-2)} \ \hat{} \, \Box_{c}  \ = \ \Box_c G
\end{equation}
We apply eq. (16) to the constant scalar $-1/\xi$
and get instead of eq. (4) now
\begin{equation}
 G^{(n+2)/(n-2)} \hat R \ = \  GR - 
\frac{1}{\xi} \Box  G
\end{equation}
We insert eq. (15) into eq. (17) and get 
\begin{equation}
\hat R =  G^{-(n+2)/(n-2)} \left(
\frac{n+2}{n-2}RG - \frac{2n}{n-2} L
\right)
\end{equation}
in place of eq. (7). The dual to eq. (18) reads
\begin{equation}
 R = \hat G^{-(n+2)/(n-2)} \left(
\frac{n+2}{n-2} \hat R \hat G - \frac{2n}{n-2} \hat L
\right)
\end{equation}
and so we get in place of eq. (9) now
\begin{equation}
\hat L =  G^{-(n+2)/(n-2)} \left(\frac{4}{n-2} R - 
\frac{n+2}{n-2}  \cdot \frac{L}{G}
\right)
\end{equation}
Applying $\frac{d}{dR}$ to eq. (18) we get
\begin{equation}
\frac {d \hat R}{dR} = 
G^{-(n+2)/(n-2)} \left[  
\frac{H(n+2)}{(n-2)^2} \left( \frac{2nL}{G} - 4R \right) - G
\right]
\end{equation}
So we suppose the r.h.s. of eq. (21) to
be unequal zero. With these conditions
the theorem of sct. 3.1. keeps valid.

\section{Discussion}

The duality operator was introduced in sct. 3.1  
to simplify the deduction of a fourth-order gravity
result, however, it seems to be applicable to other 
situations, too. 

\medskip

Let us give a non-trivial application for
fourth-order gravity (it is the last example
of subsection 3.2. but now restricted to the range $R<0$). 
We have 
$$L = \frac{3}{4} (-R)^{4/3} - \frac{1}{4},$$ 
$\hat R = 1/R$,  and $\hat L $ coincides with $L$. 
To find out all solutions with $R<0$ following from
 the corresponding field equation it suffices
to determine all solutions fulfilling $-1 \le R < 0$. 
The duality $\hat g_{ij} = (-R)^{2/3} g_{ij}$ then
gives rise to all the solutions fulfilling
$R \le -1 $, and they may be matched together
smoothly at the hypersurface
$R = -1$. So, e.g. the behaviour near a singularity 
$R \longrightarrow - \infty$ can be studied by considering the
solutions in the finite $R$-interval 
$-1 \le R < 0$. 
 Finally, it might be interesting to observe that 
the de Sitter space-time with  Hubble parameter
$h = 1/(2\sqrt 3 )$, i.e., with $R = -1$ is an
attractor solution for this field equation (at least
within the set of spatially flat Friedman models.)

\medskip 

{\it Acknowledgement}. I thank Dr. U. Kasper, Dr. M. Rainer
and one of the referees 
for making some clarifying remarks.  Financial support from the 
Wis\-sen\-schaft\-ler--Inte\-gra\-tions--Pro\-gramm 
 and from the Deutsche For\-schungs\-gemein\-schaft 
is gratefully acknowledged.

\medskip

{\large {\bf References}}

\noindent 
[1] Dobado, A. and Maroto, A. (1995) Phys. Rev. {\bf D 52},
1895.

\noindent 
[2] Rainich, G. (1925) Trans. Amer. Math. Soc. (TAMS) {\bf 27},
106.

\noindent 
[3]  Kucha\v r, K. (1963) Czech. J. Phys. {\bf B 13}, 551.

\noindent 
[4]  Bekenstein, J. (1974) Ann. Phys. NY {\bf 82}, 535.

\noindent 
[5]  Deser, S. (1984) Phys. Lett. {\bf B 134}, 419; 
Schmidt, H.-J. (1988) Phys. Lett. {\bf B 214}, 519.

\noindent 
[6] Rainer, M. (1995) Int. J. Mod. Phys. D {\bf 4}, 397;
  Rainer, M. (1995) Gravitation and Cosmology {\bf 1}, 121.

\noindent 
[7]  Bicknell, G. (1974) J. Phys. {\bf A 7}, 1061.

\noindent 
[8]  Whitt, B. (1984) Phys. Lett. {\bf B 145}, 176;
   Magnano, G., Ferraris, M. and 
 Francaviglia, M. (1987) Gen. Relat.
Grav. {\bf 19}, 465;   Schmidt, H.-J. (1987) Astron. Nachr. {\bf
308}, 183;  Barrow, J.  and  Cotsakis, S. (1988) Phys. Lett. {\bf
B 214}, 515.

\noindent 
[9]  Jakubiec, A. and  Kijowski, J. 
(1988) Phys. Rev. {\bf D 37}, 1406;  Jakubiec, A. and 
 Kijowski, J. (1989) J. Math. Phys. {\bf 30},
1073;  Jakubiec, A.  and  Kijowski, J. 
(1989) J. Math. Phys. {\bf 30}, 2923.

\noindent 
[10  Maeda, K. (1989) Phys. Rev. {\bf D 39}, 3159;  Amendola, L.,
  Litterio, M.  and  Occhionero, F. (1990) Int. J.    
Mod. Phys. {\bf A 5}, 3861.

\noindent 
[11]  Magnano, G.  and  Soko\l owski, L. 
(1994) Phys. Rev. {\bf D 50}, 5039.

\noindent 
[12]  Buchdahl, H. (1978) Int. J. theor. Phys. {\bf 17}, 149.

\noindent 
[13] Schmidt, H.-J. (1989) Class. Quant. Grav. {\bf 6}, 557.

\noindent 
[14]  Elizalde, E. and Shapiro, I. (1995) Class. Quant. Grav.
  {\bf 12}, 1385.

\noindent 
[15]   Schmidt, H.-J. (1988) Astron. Nachr. {\bf
309}, 307.

\end{document}